# Very high efficiency of low cost graphite-based solar cell by improving the fill factor using optimal ion concentration in polymer electrolyte

Dui Yanto Rahman[1], Fisca Dian Utami[1], Asep Ridwan Setiawan[2], Euis Sustini[1],

and Mikrajuddin Abdullah[1,a]

[1]Department of Physics and

[2]Faculty of Material and Aersospace Engineering

Bandung Institute of Technology

JalanGanesa 10 Bandung 40132, Indonesia

[a]Email: mikrajuddin@gmail.com

## Abstract

We report the development of graphite-based solar cells using a simple method and low cost materials. Suspension of graphite powder in mineral water was simply dropped onto the surface of fluorine-doped tin oxide glass (FTO) to form a thick film. Surprisingly, using mineral waters greatly improved the efficiency of the solar cell to reach the highest efficiency of 6.97%. Due to some minerals contained, the mineral water induced the development of fibrous structure between the graphite particles which is assumed to play a role as a bridge for the photoexcited electrons to quickly move to the electrode and suppress recombination with holes. This efficiency is very attractive when considering the materials used to develop the solar cell are all low cost. Economically this may challenge the present high efficiency semiconductor-based solar cells. We achieved the high efficiency by manipulating the cell fill factor through optimizing the ion concentration in PVA.LiOH polymer electrolyte. We also propose an equation to describe the effect of LiOH concentration and efficiency and we also provide strong correlation between the cell efficiency and the polymer conductivity.

# I. Introduction

Manufacturing of solar cell using low-cost materials, easily and readily scaled up method is demanded by the people of the world. Historically, Organic solar cell appears as a third generation since the first generation of most commercially bulk silicon-based solar cell involved expensive cost of highly needed purity material and applied technology [1]. Although inorganic thin film came afterward with a solution of using cheaper material and easier technology, it is still relatively expensive owing to high cost of encapsulation method and additionally toxic material used [1]. The most developed organic solar cell is dye-sensitized solar cell (DSSC) [2], employing dye as photon absorber, wide band gap semiconductor of $TiO_2$ as a conductive medium for the electron, and Iodine/Iodide based electrolyte as the hole transporting medium. In contrast to the first and second generation solar cells, the DSSC was made with much cheaper material and easier method. Unfortunately, it is still not readily commercial.

We have developed $TiO_2$-based solar cell using $TiO_2$ served as photon absorber [3-8] and transporting the excited electron to the electrode through metallic bridges [3]. PVA.LiOH polymer electrolyte was used as medium from transporting positively charge [3]. Considering the poor absorption of $TiO_2$ which is restricted to ultraviolet spectrum, enlarging the absorption range of $TiO_2$ remains challenging. Many attempts have been conducted to make $TiO_2$ capable of absorbing photon ranging from ultraviolet to visible region. These include the doping $TiO_2$ with various transition metal [9] and non metal such as Nitrogen (N) [10,11], carbon (C) [12-14] and boron (B), [15-17] atoms.

In previous work, we used graphite/$TiO_2$ composite in our solar cell structure as an absorbing photon material deposited with the copper particle as means of capturing the excited electron before flowing to the electrode [18]. The existence of graphite in this solar cell structure was intended to absorb a wider spectrum ranging from ultraviolet to red region, and it successfully increased the efficiency of the cell [18]. Unlike $TiO_2$, which has a wide band gap of 3.2-3.8 eV, graphite powder we used was confirmed using the calculation of tauch plot has a low band gap of 1.8 eV [18]. It is an effective band gap for active material of solar cell structure to absorb a wider range of solar light spectrum from approximately the ultraviolet to the visible region. Our graphite band gap was much higher than graphite band gap of 0.04 eV theoretically calculated by Garcia et al. [19] and 0.3 eV estimated by Vinod Kumar et al. [20]. This is due to low purity graphite powder we used [18].

In the present work, we separately use $TiO_2$ and graphite powder as a photon absorber instead of compound of them. We will investigate which of these two exhibiting better efficiency. We also used the droplet method for depositing powder film onto the transparent electrode surface. This method is much easier and more effective than the previously reported spraying method [3-8] since the spray method often finds the obstacle to its operation when the spray nozzle is frequently clogged. The electroplating process which aimed at depositing copper particles into the active material spaces to create an electron bridge at the FTO electrode was also ignored due to the decreased stability of the solar cell structure where many particles of active material were accidentally released during electroplating process. Instead, we used mineral water as dispersant liquid. The mineral water contains natural minerals extracted from the volcanic mountains. Some types of minerals contained in the mineral water are sodium, potassium, fluoride, calcium, magnesium, etc. [21]. Some of those minerals should be accidentally deposited among the graphite particles during agitation. These minerals could, hopefully, replace the function of the copper particles capturing the excited electron and allowing it to flow to the electrode.

In addition, another problem we still have in our previous work is too low fill factor belonging to our solar cell (ranging from 0.2 to 0.3) [18]. There are two factors determining the solar cell efficiency: light harvesting and fill factor [22, 23]. Light harvesting has been solved by using low band gap materials that could absorb photons in the visible even red region, while fill factor was determined by Series Resistance (Rs) and Shunt Resistance (Rp) of solar cell structure[24]. Series resistance is with respect to the junction between the interfaces of the solar cell structure [24]. Low resistance implied good junction between the interfaces. This would increase the value of the fill factor. Shunt resistance is related with current leakage due to the structure defects [24]. The fill factor increases with the shunt resistance. Increasing light absorbance region and fill factor simultaneously improve the efficiency of the cell. We are trying to solve it by increasing the ion concentration in the PVA-LiOH. Improving LiOH contents in PVA-LiOH composition is expected to increase the number of transporter of the carriers to the electrode. An increased number of the medium transport of the charge would affect the value of the fill factor. The effect of this will be confirmed by using impedance characterization.

## II. Experimental

5 g of Graphite powder (technical grade, Giva Utama, Indonesia) or $TiO_2$ powder (technical grade, Bratachem, Indonesia) was stirred homogeneously in 12 ml of mineral water (Aqua, Indonesia) for 30 minutes. FTO glass was heated on a hot plate at 200$^o$C for 30 minutes. The suspension was then poured onto the FTO surface using a small spatula. The suspension suddenly spread throughout the FTO surface. The developed film was then kept heated for 2 hours to increase good contacts between the FTO surface and the graphite particles.

Separately, 0.18 g of LiOH (Kanto, Japan) was dissolved in 20 ml of mineral water placed in a beaker for 15 minutes. Polyvinyl alcohol (PVA) (Bratachem, Indonesia) 1.8 g was then added to the solution and heated at 100 °C for 60 minutes to produce a gel-like polymer electrolyte. The solar cell module was fabricated by manually spreading the polymer electrolyte on the surface of the graphite film and fixing the aluminum electrode on the opposite side using a clamper on both sides of the cell. The schematic of the proposed solar cell is shown in **Figure_1**.

We observed morphology of graphite film using a scanning electron microscope (SEM) (JEOL JSM-6360LA, operating at 20 kV) to understand possible effect of mineral water on the film morphology. The same SEM system that has been attached with an EDX component was used to inspect the elemental compounds of the film. The X-ray diffraction (XRD) patterns were recorded using a PANalytical-X'Pert X-ray diffractometer to inspect possible occurrence of non carbon structure in the film. The performance of the solar cells was measured using a current-voltage (IV) counter (Keithley 617) under the xenon lamp illumination at intensity of 300 W / m$^2$. The illumination intensity was measured using the Newport Power / Energy Meter Model 841-PE. Electrochemical impedance spectroscopy (EIS) was recorded using a VersaSTAT 3 with a voltage amplitude of 10 mV over a frequency range of 0.05 Hz – 100 KHz. The EIS spectra were fitted by using EC-Lab V10.40 software.

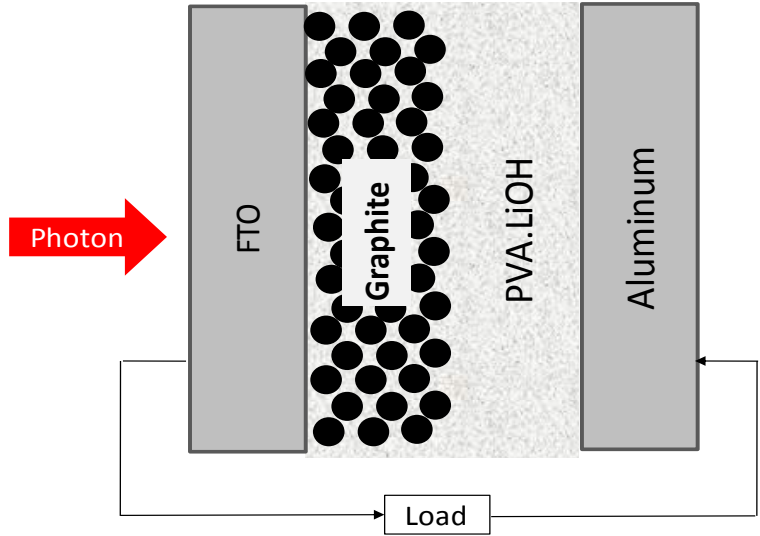

**Figure 1**.Schematic of the proposed solar cell.

## III. Results and Discussion

Initially, we tried $TiO_2$ only and graphite powder only as the active material in the solar cell for comparison. The efficiency was calculated using the equation *ε = ($P_{max}$/$P_{in}$) x 100%,* with *$P_{max}$* is the maximum value generated by multiplication of current and the corresponding voltage, and *$P_{in}$* is the power of illumination. Figure 2 shows the IV characteristic of the solar cells made of $TiO_2$ only (square) and graphite only (circle). The lines are only guide for eyes. For measurement we used the Xenon illumination intensity of 300 W/m$^2$.

We observed the solar cell made of graphite powder only shows higher efficiency than the solar cell made of $TiO_2$ powder only. The efficiency of the graphite-only solar cell was 0.76% while the efficiency of $TiO_2$–only solar cell was 0.03%. It is reasonably thought that the efficiency of solar cell containing graphite powder is much higher than that of $TiO_2$ since graphite absorbs a wider solar spectrum from the ultraviolet to the visible region as a result of the low bandgap of graphite [18]. The band gap of the graphite was observed to be 1.8 [15], while the band gap of the pure $TiO_2$ is 3.2 – 3.3 eV [25-27]. The graphite is also considered as a semimetal [28, 29].

The effectiveness of photon absorption directly affects both the $I_{sc}$ and the $V_{oc}$ of the film. We see from **Fig.2** that the graphite-based solar cell has high $I_{sc}$ (1.653 mA) and high $V_{oc}$ (0.78

V). Those values are much higher than that of the TiO$_2$-based solar cells where $I_{sc}$= 0.0847 mA and $V_{oc}$= 0.43 V.

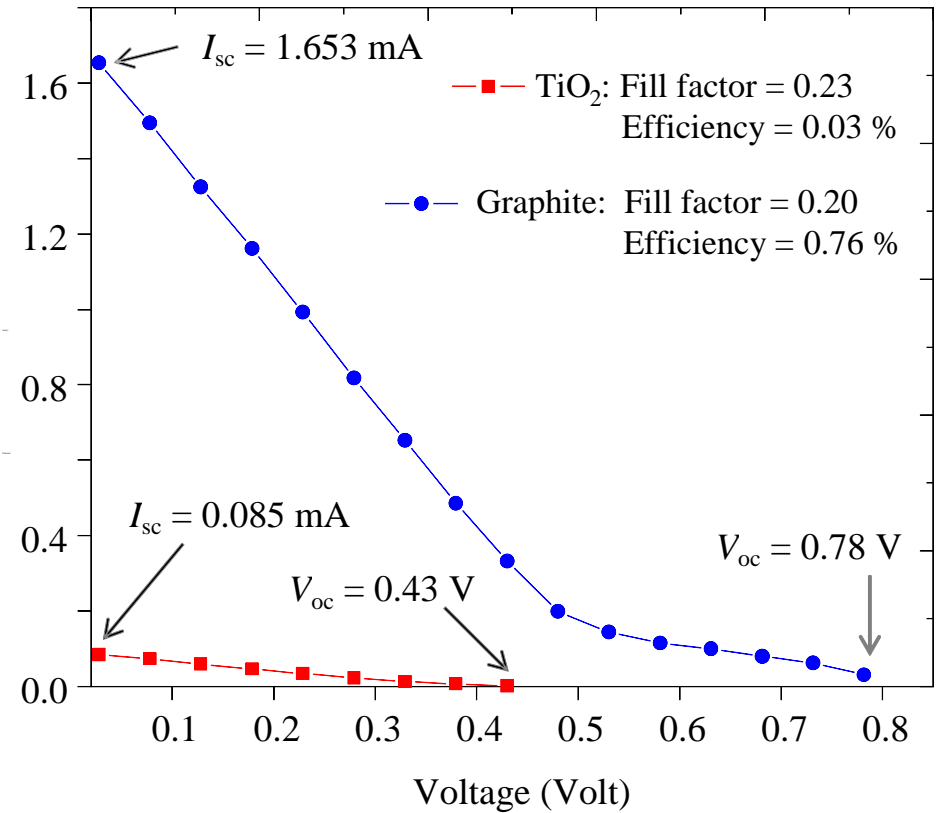

**Fig.2.** I-V characteristics of solar cell with active material of TiO$_2$ (square) and graphite powder (circle).The lines are only guide for eyes.

Interestingly, the efficiency of the solar cell made of TiO$_2$only or graphite only without copper deposition process using the mineral was not much lower than those of made with copper deposition process using aquades as reported in our previous work [18]. The efficiency of solar cell made of TiO$_2$only reported here is comparable to that of made of TiO$_2$ with copper deposition reported previously, i.e., 0.03% [18]. Both solar cells are different in basic solvent used to disperse powder, where in our previous work we used aquades [18], while in the present work we used mineral water. Similarly, the efficiency of solar cell made of graphite only was0.76%, much higher than the efficiency of solar cell made of graphite with copper deposition as reported previously as 0.12% [18].

To further understand why both methods have produced different efficiencies, we inspected the morphologies of the corresponding films prior attachment of polymer electrolyte. **Fig.3 (a)** is the graphite film made using aquades and **Fig.3(b)** is the SEM image of graphite film made using mineral water. We identified morphology differences in both films, where the film made using mineral water showed the occurrence of fibrous structures. It is likely this fibrous structure was responsible for increasing the transport of electron to the electrode, implying the reduction in the probability of electron-hole recombination. This fibrous structure plays a role as a quick route for electrons to reach the electrode. To further investigate what kind of elements that have participated to the formation of fibrous structure, we explored the elemental analysis to the graphite film made using mineral water.

**Fig.3(c)** shows the EDX spectra of the graphite film made of mineral water. We concluded that, instead of the main carbon element, some elements have been deposited in this film, namely Fe (Ferrum), Na (Natrium), P (Phosphorus), As (Arsenic), Tb(Terbium). Interestingly, those elements have not been observed in films made without mineral water (made with aquades). Some of those elements form crystalline structures as exhibited by XRD patterns as shown in **Fig.3(d)**. Inspection of the XRD patterns of the film made with mineral water showed the presence of Terbium Phosphate ($TbPO_4$), peaking $19.6°$, $34,7°$, $42,2°$, $51.2°$, $54.5°$, $61.2°$, and $65.2°$(JCPDS No. 01-075-0615), Cornwallite ($As_2Cu_5H_6O_{13}$), peaking at $19.6°$, $37.5°$, $45.1°$, and $65,2$ (JCPDS No.00-008-0162), Carbon (Graphite) peaking at $26.3°$, $42.2°$, and $54.5°$(JCPDS 00-008-0415), Tin Oxide ($SnO_2$) peaking at $26.3°$, $33.5°$, and $65.2°$ (JCPDS No.00-001-0625), and Sodium Fluoride (NaF) peaking at $37.5°$ and $54.5°$ (JCPDS No.01-075-0448).

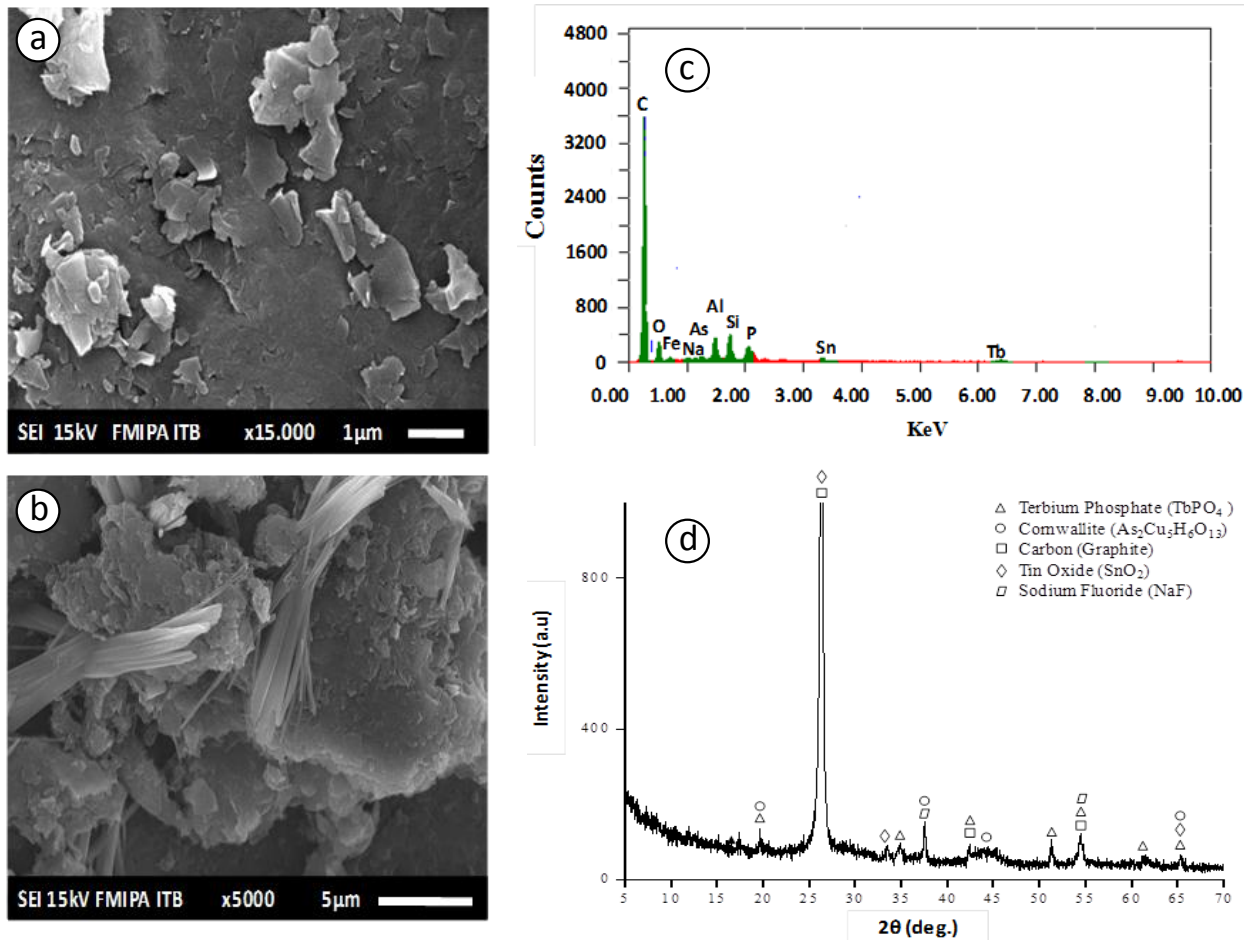

**Fig.3**(a) The graphite film made using aquades [18],(b) the SEM image of graphite film made using mineral water, (c) EDX spectra of graphite films made with mineral water, and (d) the corresponding XRD patterns.

Based on the superiority of graphite-based solar cell compared to the TiO$_2$-based solar cell, we deeply explored graphite-based solar cell by searching method for further increasing in the efficiency. The efficiency of solar cells depends on the fill factor. Based on **Fig.2**, although the $I_{sc}$ (1.653 mA) and $V_{oc}$ (0.78 V) are high enough, however, because the fill factor is too low (only 0.2) the produced efficiency becomes too low. Therefore, the possible strategy for increasing the efficiency is to improve the fill factor.

Low fill factor indicates a strong recombination of electron and hole after photo excitation. We suspect that this condition was caused by bad carrier transport of PVA-LiOH electrolyte due to low conductivity so that the positively charge ions are hardly move apart.

Therefore a possible strategy for increasing the charge transport through the electrolyte is to increase the electrolyte conductivity. It is well known that the polymer electrolyte conductivity strongly depends on the ion concentration in the electrolyte [30-34]. We have also proposed a formula for describing the effect of ion concentration on the conductivity of PVA-LiOH electrolyte [33].We calculated that the LiOH concentration of the PVA-LiOH electrolyte of 9% as used in the above mentioned graphite-based solar cell is still low and theoretically, the maximum conductivity occurs in the range of ion concentration from 15% to 20% [33, 34]. Therefore, we tried to fabricate the solar cell using high LiOH concentrations, ranging from 9% to 25%.

**Fig.4** shows effect of LiOH concentration on the efficiency (diamond symbols) and fill factor (square symbols) of graphite-based solar cell. Curves are merely for eye guidance. The values in parentheses represents ($I_{sc}$,$V_{oc}$). We observed the fill factor of only 0.18 when using LiOH concentration of 9% increased to 0.28 when using LiOH concentration of 13%. The increase in the fill factor was followed by a significant increase in the efficiency of the solar cells from 0.76% when using LiOH concentration of 9% to become 2.59% when using LiOH concentration of 13%. The main factor that is responsible for increasing the fill factor is the increase in $I_{sc}$, since $V_{oc}$ nearly unchanged in two conditions. It is likely caused by more electron and holes have been transported to produce high current. The increase in the fill factor continues by the increase in the LiOH concentration up to 17%, which is likely the optimal condition. Further increase in the LiOH concentration from 21% to 25% reduces back the fill factor.

The highest observed efficiency of 6.97% was achieved with a fill factor of 0.48. It is obtained at LiOH concentration of 17% and at this condition, the measured $I_{sc}$ = 3.66 mA and $V_{oc}$= 1.18 V. An interesting phenomenon was observed related to the decrease of the cell efficiency after the increase in the LiOH concentration to more than 17% although $I_{sc}$ and $V_{oc}$ were still high. For a more complete explanation of this phenomenon, we performed an EIS characterization to understand how the polymer resistance changes with LiOH concentration.

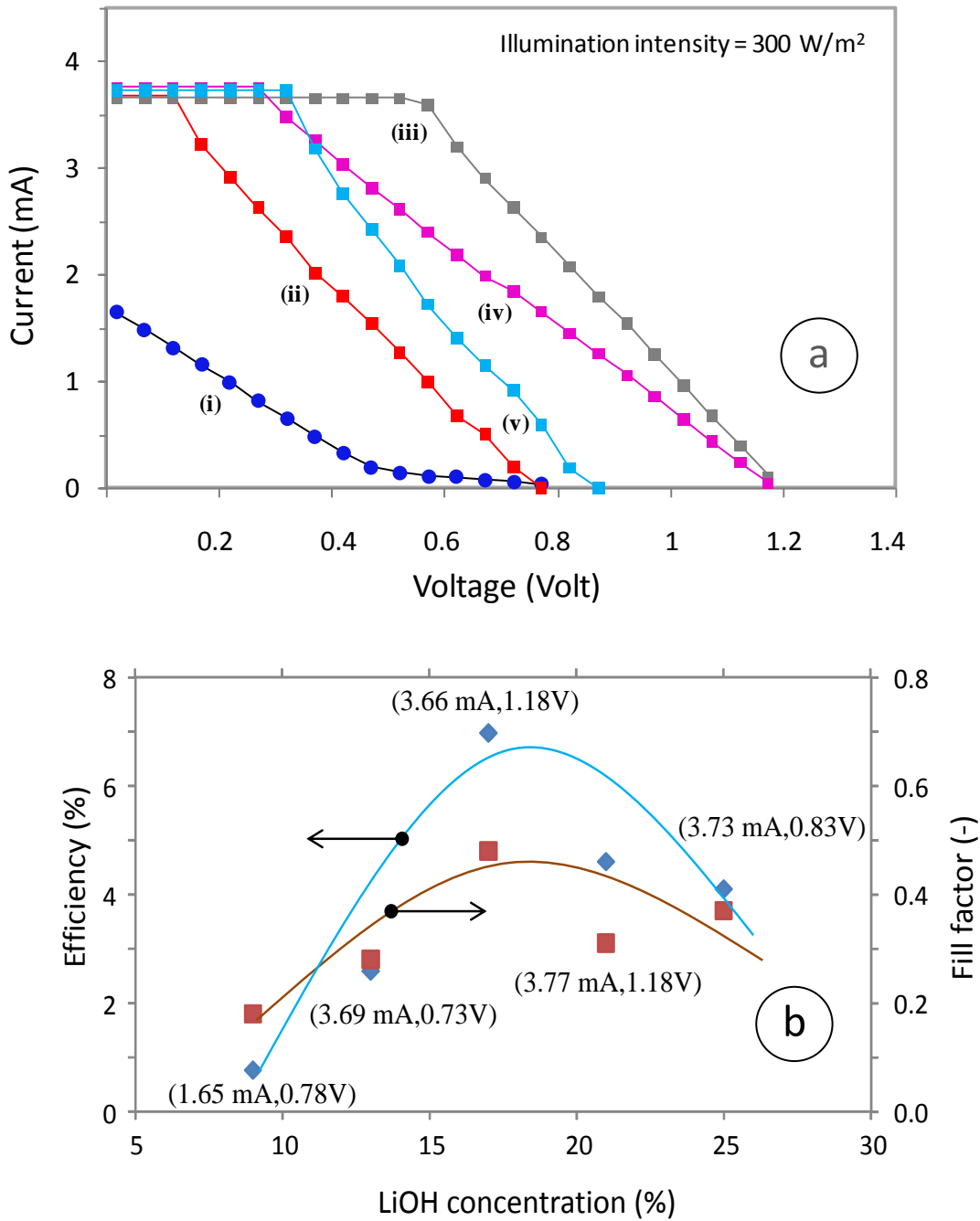

**Fig.4.** I-V (a) characteristics with variation of LiOH concentration in PVA-LiOH polymer electrolyte (i) 9% (ii) 13% (iii) 17% (iv) 21% and (v) 25% (b) performance parameters, of solar cell with active material of graphite powder.

Based on the above findings, we can summarize the process of efficiency enhancement as illustrated in **Fig.5**. **Fig.5(a)** shows the condition when the graphite film was prepared without mineral water and the conductivity of the electrolyte was low due to low LiOH concentration. The fibrous structure was absent in the space between graphite particles. The negatively and positively charges are hardly move to the opposite direction so that the probability of recombination is high to result low efficiency. **Fig.5(b)** shows the condition when the graphite film was prepared with mineral water but the conductivity of the electrolyte was low due to low LiOH concentration. The electrons have bridges to move quickly to the electrode while the positively charges are still hard to move to the opposite direction due to low polymer conductivity. Slightly enhancement in the efficiency is observed, compared to condition in Fig. 8(a). Finally, in **Fig. 5(c),** the mineral water was used to deposit graphite and the concentration of LiOH used in the polymer was optimum. Electron can move quickly to the electrode due to the presence of bridges and positively charges can transport quickly to the opposite direction due to high polymer conductivity. The total result is high enhancement in efficiency.

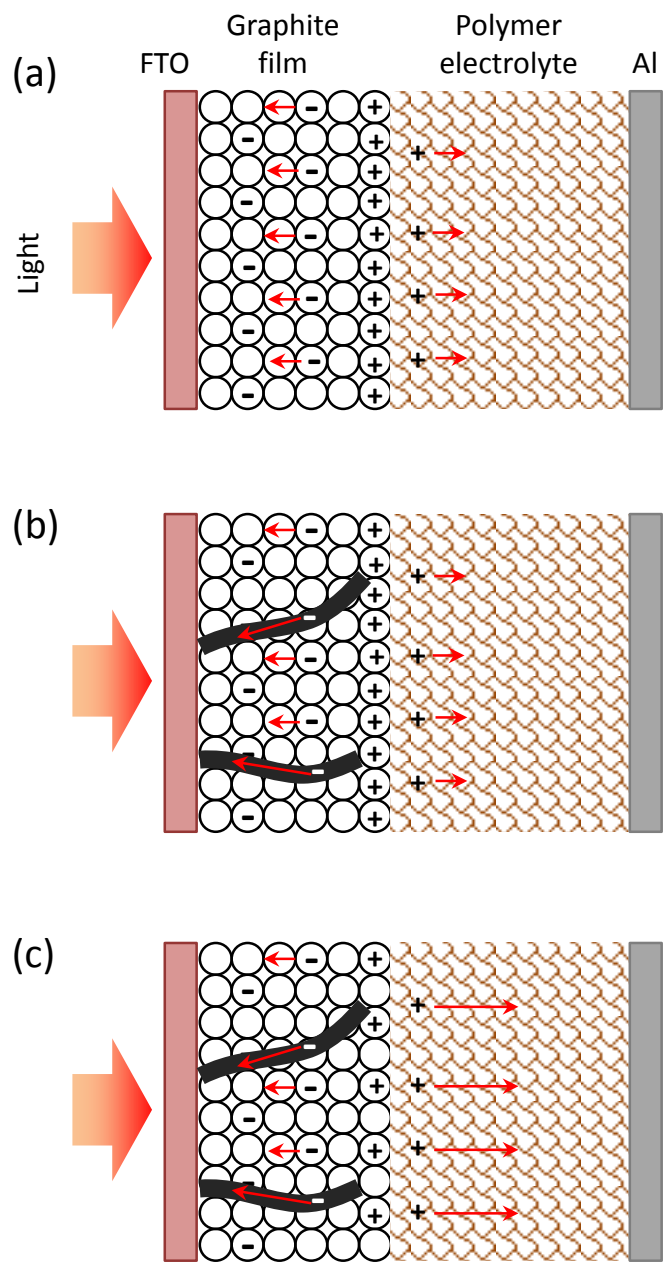

**Fig.5.** The Illustration of efficiency enhancement due to the use of mineral water and optimum LiOH concentration. (a) The graphite film was prepared without mineral water and the conductivity of the electrolyte was low due to low LiOH concentration. Positively and negatively charges hardly move to the opposite direction. (b) The graphite film was prepared with mineral water but the conductivity of the electrolyte was low due to low LiOH concentration. Negatively charges can move quickly while the positively charges hardly move to the opposite direction. (c)

The mineral water was used to deposit graphite and the concentration of LiOH used in the polymer was optimum. Positively and negatively charges can move to the opposite direction.

By comparing the change of efficiency of the solar cell with respect to LiOH concentration we concluded that this figure nearly the same as the change of electrical conductivity of the polymer electrolyte with respect to LiOH concentration as we reported previously [33]. Therefore the equation that describes the dependence of efficiency on the LiOH concentration must be similar to that of electrical conductivity with respect to LiOH. We have derived such the equation for electrical conductivity [33] and when the same equation is applied to efficiency we obtain the following approximated equation

$$\varepsilon = \varepsilon_{max} e^{-\alpha(1/C^{1/3} - 1/C_0^{1/3})^2} \tag{1}$$

Where $\varepsilon$ in %, $\varepsilon_{max}$ is the maximum efficiency that occurs at $C_0$, and $\alpha$ are constants, C is the salt concentration (in %) and $C_0$ is the concentration (in %) when the efficiency is maximum. Eq. (1) can be rewritten as $\ln \varepsilon = \ln \varepsilon_{max} - \alpha(1/C^{1/3} - 1/C_0^{1/3})^2$. By defining $x = 1/C^{1/3}$ we obtained the following quadratic equation

$$\ln \varepsilon = \ln \varepsilon_{max} - \alpha(x - x_0)^2 \tag{2}$$

To obtain the parameter $\varepsilon_{max}$ and $\alpha$ we fit the measured data with aquadratic equation. **Figure 6** shows the fitting result. We obtained the best fitting equation as $\ln \varepsilon = -179.12x^2 + 134.56x - 23.623$ with $R^2 = 0.9367$. Based on this fitting result we obtain $\alpha = 179.12$ and $2\alpha x_0 = 134.56$, resulting $x_0 = 134.26/2\alpha = 0.3756$. Since $x_0 = 1/C_0^{1/3}$ we then have $C_0 = 18.87$ %.

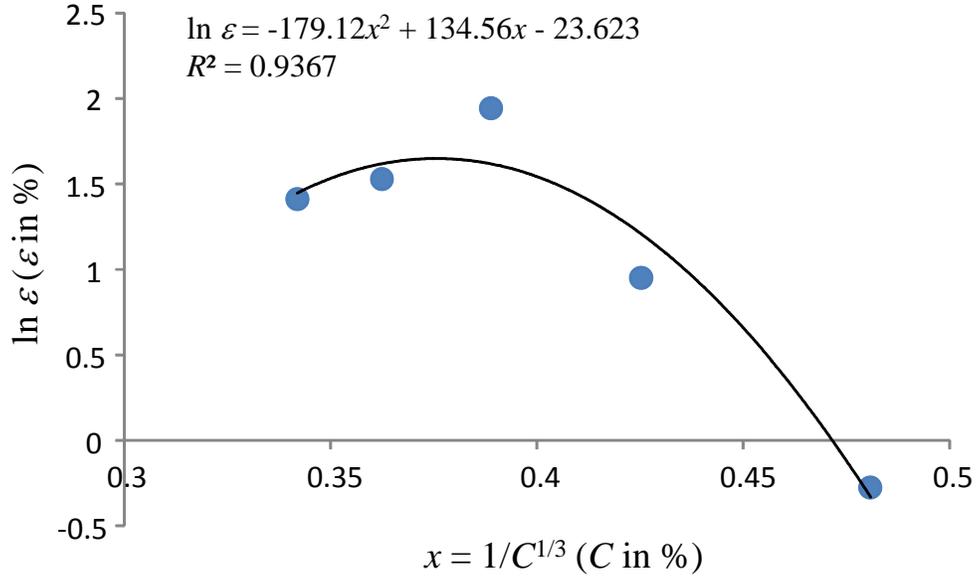

**Fig.6**. Effect of LiOH concentration, expressed by variable $x = 1/C^{1/3}$ with C is the LiOH concentration on the efficiency of solar cell. Circles are the measured data and curve is the fitting curve.

By inspecting **Fig.4** we also conclude that the fill factor changes with respect to LiOH concentration similar to the efficiency so that the fill factor also satisfies similar equation. Therefore, the fill factor might be approximated as

$$FF = FF_{max} e^{-\alpha(1/C^{1/3} - 1/C_0^{1/3})^2} \tag{3}$$

With $FF$ is the fill factor and $FF_{max}$ is the maximum fill factor that occurs at $C_0$.

**Fig.7** (a) shows the result of the EIS measurement and (b) the corresponding equivalent circuit. The fitted EIS parameters are summarized in **Table 1**. Generally, the Nyquist curves are composed of three semicircles. The first high-frequency semicircle describes the charge transfer resistance between the polymer electrolyte and the aluminum counter-electrode ($Rd_4$). The second semicircle, in the middle, is associated with the charge transfer resistance at the photon absorber material/electrolyte interface ($R_2 + R_3 + R_4$). The third semicircle represents the resistance of the FTO /photon absorber material contact ($R_1$)

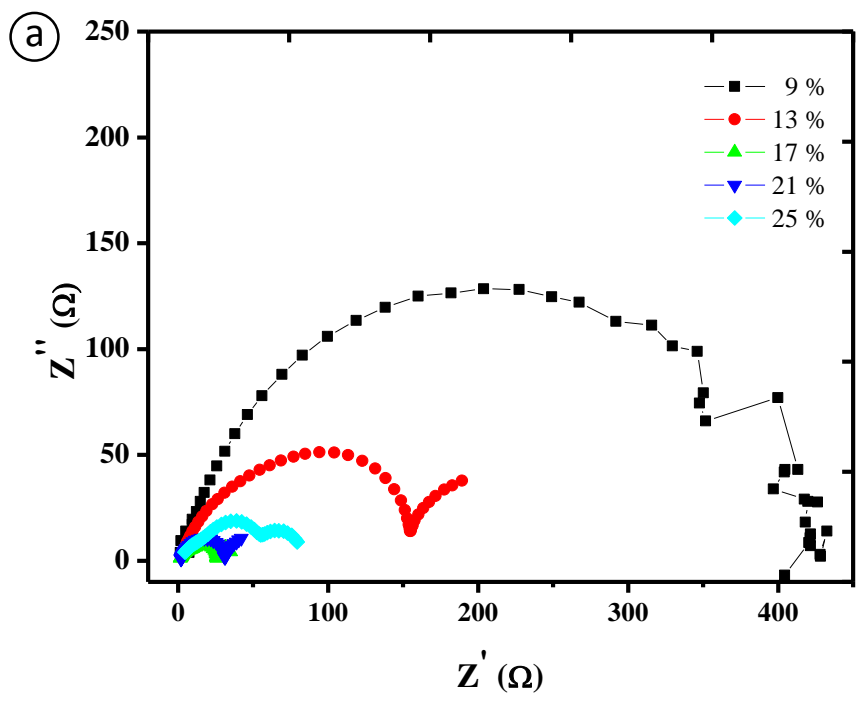

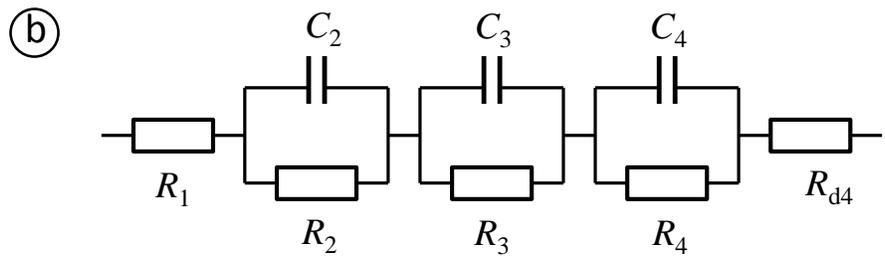

**Figure 7**(a) Nyquist plot of solar cell impedance at different LiOH concentrations in PVA-LiOH polymer electrolyte and (b) the corresponding equivalent circuit.

**Table 1**.

Summary of fitted parameters of EIS result

| LiOH Concentration (%) | $R_1$ (Ohm) | $R_2$ (Ohm) | $R_3$ (Ohm) | $R_4$ (Ohm) | $R_2+R_3+R_4$ (Ohm) | $Rd_4$ (Ohm) |
|---|---|---|---|---|---|---|
| 9  | 2.971 | 164.2 | 203   | 42.99 | 410.19 | 152.9 |
| 13 | 0.651 | 18.9  | 71.55 | 82.34 | 172.74 | 53.22 |
| 17 | 0.896 | 5.01  | 1.343 | 10.71 | 17.06  | 18.57 |
| 21 | 1.703 | 1.18  | 7.31  | 18.82 | 27.31  | 22.51 |
| 25 | 2.798 | 5.918 | 30.39 | 9.798 | 46.11  | 32.56 |

The resistance of the FTO/photon absorber ($R_1$) contact appears to be relatively constant, ranging from 1 to 3 ohms. Although there are a few changes with an increase in the concentration of LiOH which is probably caused by the polymer electrolyte penetration into the graphite particle film spaces. The information explaining the increase and decrease in the efficiency of the solar cells will be clearly observed in the change of the charge transfer resistance of the graphite / electrolyte interface ($R_2 + R_3 + R_4$).

We have assumed that the increase in efficiency was caused by increase in the polymer conductivity. The data in **Table 1** is the resistance, and the polymer resistance is represented by $R_2+R_2+R_4$. For polymer with fixed dimension we have $R_2 + R_3 + R_4 \propto \sigma$, with σ is the polymer conductivity. Therefore, we can write the equation for LiOH concentration dependence of polymer resistance as

$$(R_2 + R_3 + R_4) = (R_2 + R_3 + R_4)_{max} \, e^{\alpha(1/C^{1/3} - 1/C_0^{1/3})^2} \tag{4}$$

In the logarithmic form we can write

$$\ln(R_2 + R_3 + R_4) = \ln(R_2 + R_3 + R_4)_{max} + \alpha(x - x_0)^2 \tag{5}$$

**Fig.8** shows the effect of LiOH concentration on the ($R_2 + R_3 + R_4$). Again we fit the data with a quadratic function and we obtain the best fitting equation as $\ln(R_2 + R_3 + R_4) = 178x^2 - 126.78x + 26.033$ with $R^2 = 0.7782$. From this resulted equation we obtain $\alpha = 178$ and $C_0 = 22.14$ %. Surprisingly the obtained parameter α in both fittings (fitting of efficiency and fitting of resistance) is exactly the same, and the obtained parameter $C_0$ is nearly the same. This result supports the hypothesis that the efficiency and the resistance of conductivity of the polymer are strongly correlated and the increase in the efficiency was caused by increase in the polymer conductivity.

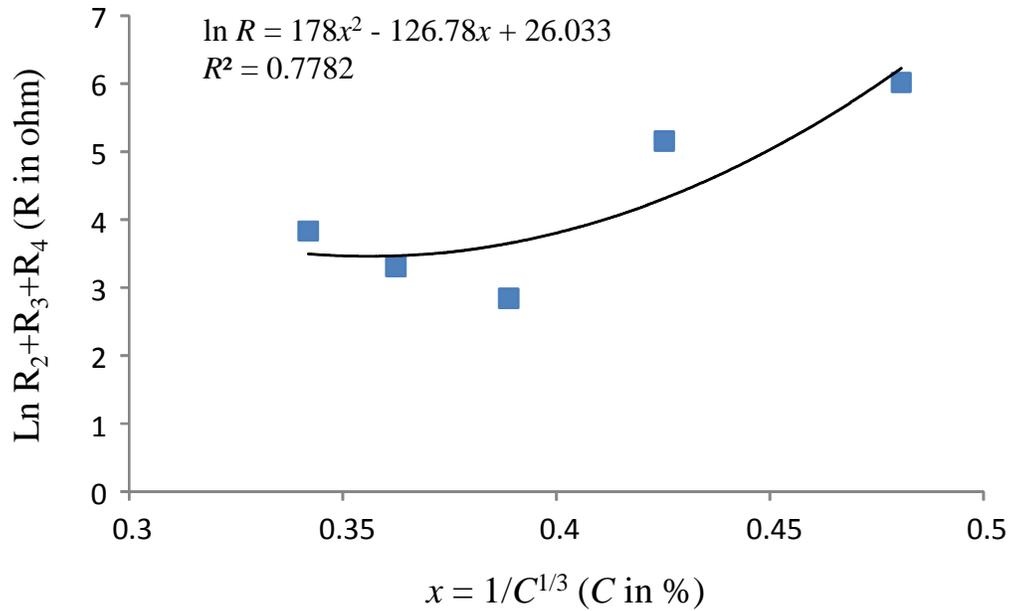

**Fig.8.** Effect of LiOH concentration, expressed by variable $x = 1/C^{1/3}$ with $C$ is the LiOH concentration on the sum $R_2+R_3+R_4$. Squares are the measured data and curve is the fitting curve. The sum $R_2+R_3+R_4$ determines the resistance of the polymer electrolyte.

The sufficient amount of $Li^+$ will be very beneficial for transporting the holes of electron-holes produced from active material, and simultaneously reduce electron-hole recombination. The $Li^+$ concentration has optimal value (in this work is about 17% w/w), associated with optimal distance between the ions to move easily throughout the polymer and reach the counter electrode [33, 34]. There will be repulsion between ions if the distance is too short that will inhibit the ions movement to counter electrode [33,34]. A LiOH concentration greater than17% make the distance between the ions too short which implied to inhibited movement of ions to the counter electrode despite numerous electron-hole pairs produced during the absorption of the photons by the active material. Meanwhile, the concentration of LiOH below 17% corresponds to the distance of ions that is too far away. A distance too far apart will involve a long distance ion jump during its movement throughout the polymer. Additionally, the $R_2 + R_3+R_4$ behavior in explaining the efficiency of solar cell was supported by the same behavior of the charge transfer resistance between the polymer electrolyte and the aluminum counter-electrode ($Rd_4$).

## IV. Conclusion

We have successfully fabricated a new solar cell using graphite powder as a photon absorber. We obtained a high efficiency up to 6.97%, a very attractive value for solar cells made using low cost materials and simple preparation method. Equation describes the relationship of the solar cell efficiency with the ion concentration in polymer was also proposed. This solar cell design was promising for future production on a larger scale owing its very easy and cheap manufacturing process.

## Acknowledgment

This work was supported by fellowship grand from Indonesia Endowment Fund for Education (LPDP)

[3] S. Saehana, P. Arifin, Khairurrijal, M. Abdullah, A new architecture for solar cells involving a metal bridge deposited between active TiO 2 particles, J. Appl. Phys. 111 (2012). doi:10.1063/1.4730393.

[4] S. Saehana, R. Prasetyowati, M.I. Hidayat, P. Arifin, Efficiency Improvement in TiO2-Particle Based Solar Cells after deposition of metal in spaces between particles, Int. J. Basic Appl. Sci. 1 (2011) 15–28.

[5] S. Saehana, Darsikin, E. Yuliza, P. Arifin, Khairurrijal, M. Abdullah, A new approach for fabricating low cost DSSC by using carbon-ink from inkjet printer and its improvement efficiency by depositing metal bridge between titanium dioxide particles, J. Sol. Energy Eng. Trans. ASME. 136 (2014). doi:10.1115/1.4027695.

[6] M. Rokhmat, E. Wibowo, Sutisna, E. Yuliza, Khairurrijal, M. Abdullah, Enhancement of TiO2 Particles Based-Solar Cells Efficiency by Addition of Copper(II) Nitrate and Post-Treatment with Sodium Hydroxyde, Adv. Mater. Res. 1112 (2015) 245–250. doi:10.4028

[7] E. Yuliza, S. Saehana, D.Y. Rahman, M. Rosi, Khairurrijal, M. Abdullah, Enhancement performance of dye-sensitized solar cells from black rice as dye and black ink as counter electrode with inserting copper on the space between TiO2 particles by using electroplating method, Materials Science Forum (Vol. 737, pp. 85-92). Trans Tech Publications. (2013). doi:10.4028

[8] M. Rokhmat, Sutisna, E. Wibowo, Khairurrijal, M. Abdullah, Efficiency enhancement of TiO2(active material) solar cell by inserting copper particles grown with pulse voltage electroplating method, J. Phys. Chem. Solids. 100 (2017) 92–100. doi:10.1016/j.jpcs.2016.09.019.
19